# SABR Type Libor (Forward) Market Model (SABR/LMM) with time-dependent skew and smile

Osamu Tsuchiya

(Independent)

Abstract

Volatility Skew and Smile of Interest Rate products (Swaption and Caplet) are represented by SABR (Stochastic Alpha Beta Rho model). So, the Interest Rate derivatives model for pricing the callable exotic swaps should be comparable to the SABR volatility surface. In the interest rate derivatives models, Libor Market Model (LMM) (in a post-Libor world, Forward Market Model (FMM)) is one of the most popular models used in the market. So, there are many attempts to develop LMMs that are comparable to the SABR surface. It is called SABR/LMM. There are many references for SABR/LMM, but most of them only treat SABR/LMM, which is not flexible enough to be used practically in global banks. The purpose of this paper is to provide a comprehensive definition of SABR/LMM and a complete description of how it is to be implemented.

Introduction

Libor Market Model (LMM) (Forward Market Model (FMM) in post-Libor Risk-Free Rate) is a multi-factor interest rate model, which is widely used for the valuation of callable interest rate swaps[1]. It has the following properties[1][8][9].

1. Arbitrage freeness is guaranteed by drift
2. Volatility structure is flexible enough to be fitted to all of the components of the Swaption Volatility matrix
3. It has enough correlation structure to be fitted to spread option market
4. Rates dynamics are transparent

---

[1] Other models which are based on short rate such as the Hull-White model [6] or Cheyette [1] are used in the market also.

5. It is possible to include Local and Stochastic Volatilities to explain the skew and the smile of the volatility surface

Most callable exotic interest rate swaps are hedged by the swap and swaption (and sometimes spread option) market, so the model volatilities should be calibrated to swaption volatilities. (Here caplet is considered as a swaption with minimum length. In the RFR backward-looking rate adjustment from the time decay of volatility is necessary.)[11]

SABR and Uncorrelated SABR

In the current market standard, volatility skew and smile of Swaption (and Caplet) is represented by SABR parameters α(0), β, ρ and ν. In SABR, implied volatility is a function of SABR parameters
$$\sigma_{Hgan}(\alpha(0), \beta, \rho, \nu) \quad (1)$$
Here the function is an approximation for the implied volatility of the SABR model
$$dF(t) = \alpha(t)F^{\beta}(t)dW(t) \quad (2)$$
$$d\alpha(t) = \nu\alpha(t)dZ(t)$$
$$dW(t)dZ(t) = \rho dt$$
which was derived by Hagan et. Al [13]. The approximation is based on singular perturbation expansion, and the approximation is generally fine but for long maturity or high volatility environment, the error can be larger. Because an approximation formula represents the volatility surface, even if our term structure model fits the SABR model exactly, there can be a large difference in volatility surface between the term structure model and the Hagan approximation represented by SABR parameters.

On the other hand, in the SABR model, when the correlation between rates and volatility is zero, the exact solution is available (Antonov et. al.[12]). And in the SABR parameters, correlation $\rho$ has a similar role as $\beta$. That is correlation parameter can be considered redundant. So, we can use the SABR model with the correlation being zero to represent the volatility surface.

In this situation, volatility surfaces of Swaptions (and Caplet) are assumed to be represented by the exact solution of the SABR model with the correlation between rates and volatility being zero and this is called **Uncorrelated SABR**.

$$dS(t) = \alpha(t)\left(\frac{S(t)}{S(0)}\right)^{\beta} dW(t) \quad (9) \quad d\alpha(t) = v\alpha(t)dZ(t)$$

$$dW(t)dZ(t) = 0$$

Note that definition of $\alpha(0)$ in our formulation is different to market standard one by a factor $S(0)^{\beta}$ and Swap Rate and Stochastic Volatility are assumed to be independent each other.

Fig 1.

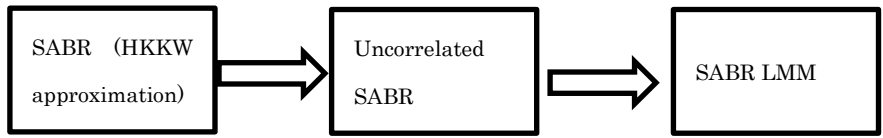

### Volatility Skew and Smile

There are many types of models that explain volatility skew and smile. In this paper, skew and smile are explained as followings. Volatility skew is explained by local volatility like the CEV model or the Displaced Diffusion model. On the other hand, the smile is explained by the stochastic volatility model. In this respect, we adopt the following SABR. The skew is explained by the CEV type local volatility model and the control parameter is $\beta$. The smile is represented by the Stochastic Volatility model and the control parameter is $v$. The correlation between interest rate and stochastic volatility $\rho$ has almost the same role aa s $\beta$ for the representation of volatility surface and it is redundant. So we assume $\rho = 0$. In this case, option valuation is given by exact formula (AKRS), where in the SABR we can use only asymptotic expansion.

### Correlation of yield curve

Many exotic interest rate derivatives depend on the correlation between the term structure of the yield curve [1][5]. The most notable example is CMS spread swap. So, the power of Libor Market Model is most apparent for the pricing of the callable derivatives whose underling is a CMS spread. For the pricing of these products, the model should be calibrated to CMS spread option market.

## Model Definition

### Notation

We assume cash flows of the interest rate derivatives we are interested in are on the tenor structure

$$[0 = T_0, T_1, \cdots, T_N, T_{N+1}]$$

Forward Libor for the period $[T_j, T_{j+1}]$ observed at t is $L_j(t)$. The day count fraction is $\delta_j = T_{j+1} - T_j$. Discount factor with maturity $T_j$ observed at t is given by $D(t, T_j)$.

After the LIBRO replacement, backward looking rate $R_j(T_{j+1})$ that will be paid at $T_{j+1}$ is a compounding average of the over night rate for the period $[T_j, T_{j+1}]$ that is $\delta_j R_j(T_{j+1}) = \prod_k^n(1 + r_k(t_k))$, where $T_j = t_1, \cdots, t_n = T_{j+1}$, $r_k(t_k)$ is an overnight rate for the period $[t_k, t_{k+1}]$. The value of backward looking rate before $T_j$ is represented by the same formula as forward Libor, and the valuation of most Swaptions are given in the same framework as LMM. Regarding Caplet, the decay of volatility needs to be treated [10] but also in the same framework as in the LMM. Only for the valuation of the special products for example Tarns need actual simulation of the backward looking rate.

Swap rate, which has interest exchanges at $T_{n+1}, T_{n+2}, \cdots, T_{m+1}$ is denoted by $S_{nm}(t)$. Where there are not any confusion, we write it as $S(t)$. The swaption which expire at $T_n$ and enter into swap with the above exchange dates are also represented by $(n, m)$.

The swap rate is given by

$$S_{nm}(t) = \frac{\sum_{j=n}^m \delta_j^{flt} L_j(t) D(t, T_{j+1})}{Ann_{nm}(t)} \quad (3)$$

where

$$Ann_{nm}(t) = \sum_{j=n}^m \delta_j^{fix} D(t, T_{j+1}) \quad (4)$$

In this paper, volatility surface of the swaption with underlying is this swap and exercise date is $T_n$ will be analyzed.

Here some of the simplification is made for both calibration instruments and Monte Carlo pricing. Floating leg and fixed leg of the swap is assumed to have the same payment schedule, but they can have different day count fractions.

The Libor schedule is a three-dates schedule that is Libor payment date is equal to the end of the Libor calculation period.

## SABR/LMM

SABR/LMM is given by a stochastic differential equation of the Libor rates

$$dL_i(t) = \mu_i(t)dt + \alpha_i(t)g_i(t)\phi_i(t, L_i(t))dW_i(t) \quad (5)$$
$$d\alpha_i(t) = \nu_i(t)\alpha_i(t)dZ_i(t)$$

Where

$$dW_i(t)dW_j(t) = \rho_{ij}(t)dt$$
$$dW_i(t)dZ_j(t) = 0$$
$$dZ_i(t)dZ_j(t) = r_{ij}(t)dt$$
$$\alpha_i(0) = 1$$

The correlation structure of rates has the following form [5]

$$\rho_{ij}(t) = \exp(-\beta(t)|T_i - T_j|) \quad (6)$$

The Brownian motion of rates $dW_i(t)$ and stochastic volatilities $dZ_j(t)$ are uncorrelated. Such that the stochastic volatilities processes are driftless. Here there is not market instruments to calibrate the correlations between Stochastic Volatilities, so we can choose them being perfectly correlated $r_{ij}(t) = 1$.

Regarding local volatility, we can adopt both CEV (SABR) type and Displaced Diffusion (DD) type Local Volatilities

In CEV

$$\phi_i(t, L_i(t)) = \left(\frac{L_i(t)}{L_i(0)}\right)^{\beta_i(t)} \quad (7)$$

In DD

$$\phi_i(t, L_i(t)) = \frac{L_i(0) + \beta_i(t)(L_i(t) - L_i(0))}{L_i(0)} \quad (8)$$

Our approximation formula used in the calibration is common for both CEV and DD. The distribution of Monte Carlo paths is of course different (especially in the low rate area).

All of the parameters in our SABR/LMM are assumed to be piecewise constant in observation time between tenor dates.

## Approximation formula for Swap Rate SABR parameters

Let uncorrelated SABR dynamics of Swap Rate $S(t)$ be given by

$$dS(t) = \alpha(t)\left(\frac{S(t)}{S(0)}\right)^{\beta} dW(t) \quad (9)$$

$$d\alpha(t) = \nu\alpha(t)dZ(t)$$

$$dW(t)dZ(t) = 0$$

The parameters are assumed to fit to market volatility by an exact formula given in reference [12]. In the market, swaption volatility is generally represented by usual SABR parameters (including ρ), and uncorrelated SABR should be fitted to it before calibration of SABR/LMM.

Before deriving approximation formula for Swap Rate SABR, we investigate Swap Rate dynamics in SABR/LMM. In Swap measure, Swap Rate dynamics is given by

$$dS(t) = \sum_{i=n}^{m} \frac{\partial S(t)}{\partial L_i(t)} dL_i(t) = \sum_{i=n}^{m} v_i(t)\alpha_i(t)g_i(t)\phi_i(t, L_i(t))dW_i(t)$$

$$= \lambda(t, \{L\}(t), \{\alpha\}(t))dW(t) \quad (10)$$

where

$$\lambda^2(t, \{L\}(t), \{\alpha\}(t)) = \sum_{i,j=n}^{m} v_i(t)v_j(t)\alpha_i(t)\alpha_j(t)g_i(t)g_j(t)\phi_i(t, L_i(t))\phi_j(t, L_j(t))\rho_{ij}(t) \quad (11)$$

Here drift term should be 0 in Swap measure and $v_i(t) = \partial S(t)/\partial L_i(t)$. Actually, Swap Rate depends on Libor rates $L_i(t)$ with $i = 0, \cdots, m$ but here we assume it depends on the Libor with $i = n, \cdots, m$ only. Here we adopt simple approximation

$$v_i(t) = \frac{\delta_i D(t, T_{i+1})}{Ann(t)} \quad (12)$$

And we assume also

$$\frac{\partial v_i(t)}{\partial L_i(t)} = 0$$

Note that on the ATM paths $(L_i(t) = L_i(0)$ for all $i)$, $v_i(t) = v_i(0)$

ATM Approximation for Skew Parameter $\beta$

Before mapping SABR/LMM to uncorrelated SABR, we will map SABR/LMM to Time Dependent SABR

$$dX(t) = \alpha_X(t) g_X(t) \left(\frac{X(t)}{X(0)}\right)^{\beta_X(t)} dW(t) \quad (13)$$

$$d\alpha_X(t) = \nu_X(t) \alpha_X(t) dZ(t)$$

$$\alpha_X(0) = 1$$

Note that here all of the parameters (volatility parameter, skew parameter and volatility of volatility) are time dependent.

Now time dependent volatility parameter $g_X(t)$ and stochastic volatility process $\alpha_X(t)$ are determined such that they fit to SABR/LMM on the ATM path

$$g_X^2(t) = \sum_{ij=n}^{m} v_i(0) v_j(0) g_i(t) g_j(t) \rho_{ij}(t) \quad (14)$$

$$\alpha_X^2(t) = \left(\frac{1}{g_X^2(t)}\right) \sum_{ij=n}^{m} v_i(t) v_j(t) \alpha_i(t) \alpha_j(t) g_i(t) g_j(t) \rho_{ij}(t) \quad (15)$$

This assumption makes time dependent SABR match to Swap Rate volatility (1) on ATM path.

Now we will search $\beta_X(t)$ which makes slope of variance for time dependent SABR match that of variance for Swap Rate dynamics in SABR/LMM [1].

A derivative of variance in Swap Rate process in (11) with respect to $L_i(t)$ at ATM and at the initial stochastic volatility is given by

$$\left.\frac{\partial \lambda^2(t, \{L\}(t), \{\alpha\}(t))}{\partial L_i(t)}\right|_{ATM} = \frac{2\beta_i(t)}{L_i(0)} v_i(0) \alpha_i(t) g_i(t) \sum_{j=n}^{m} v_j(0) \alpha_j(t) g_j(t) \rho_{ij}(t) \quad (16)$$

On the other hand, a derivative of variance in the time dependent SABR model (13) at ATM is given by

$$\left.\frac{\partial}{\partial L_i(t)} \alpha_X^2(t) g_X^2(t) \left(\frac{S(t)}{S(0)}\right)^{2\beta_X(t)}\right|_{ATM} = 2\alpha_X^2(t) \beta_X(t) \frac{v_i(0)}{S(0)} \quad (17)$$

There are $m - n + 1$ relationships corresponding to $i = n, \cdots, m$. To have best fit $\beta_X(t)$, we need to solve optimization problem

$$\min_{\beta_X(t)} f(\beta_X(t))$$

$$f(\beta_X(t)) = \sum_{i=n}^{m} \left\{ \frac{2\beta_i(t)}{L_i(0)} v_i(0) \alpha_i(t) g_i(t) \sum_{j=n}^{m} v_j(0) \alpha_j(t) g_j(t) \rho_{ij}(t) - 2\alpha_X^2(t) \beta_X(t) \frac{v_i(0)}{S(0)} \right\} \quad (18)$$

This is quite a complex equation, but by omitting common factor $v_j(0)$ (which will not affect the result so much), we have simpler problem and solution of it is given by

$$\beta_X(t) = \frac{S(0)}{\alpha_X^2(t)g_X^2(t)} \sum_{j=n}^{m} \frac{\beta_j(t)}{L_j(0)} \alpha_j(t)g_j(t) \sum_{i=n}^{m} v_i(0)\alpha_i(t)g_i(t)\rho_{ij}(t) \quad (19)$$

Now we have a mapping from SABR/LMM to time dependent SABR.

Skew Averaging

In this section, we will derive the local stochastic volatility model with time independent skew parameter (SABR)

$$dY(t) = \alpha_X(t)g_X(t)\left(\frac{Y(t)}{Y(0)}\right)^B dW(t) \quad (20)$$

Which has similar plain vanilla option prices for maturity $T_n$ as the one with time dependent SABR,

$$dX(t) = \alpha_X(t)g_X(t)\left(\frac{X(t)}{X(0)}\right)^{\beta_X(t)} dW(t) \quad (21)$$

Here we assume stochastic volatility process is the same in the above two processes.

The key technique used in this subsection is a parameter average (skew average) [1] [3]. In the skew averaging technique, we search the time independent local volatility model which has the closest option price with the time dependent local volatility model near ATM region.

Let $h(t, x)$ be a time dependent and $\bar{h}(x)$ a time independent local volatility functions

$$h(t, x) = \left(\frac{x}{X(0)}\right)^{\beta_X(t)} \quad (22)$$

$$\bar{h}(x) = \left(\frac{x}{X(0)}\right)^B \quad (23)$$

To analyze the model in the region close to ATM, we define rescaled local volatilities

$$h_\epsilon(t, x) = \left(\frac{X(0) + (x - X(0))\epsilon}{X(0)}\right)^{\beta_X(t)} \quad (24)$$

$$\bar{h}_\epsilon(x) = \left(\frac{X(0) + (x - X(0))\epsilon}{X(0)}\right)^B \quad (25)$$

And define two families of stochastic processes indexed by $\epsilon$ corresponding to the above processes

$$dX_\epsilon(t) = \alpha_X(t)g_X(t)h_\epsilon(t, X_\epsilon(t))dW(t) \quad (26)$$
$$dY_\epsilon(t) = \alpha_X(t)g_X(t)\bar{h}(Y_\epsilon(t))dW(t) \quad (27)$$

Now we will define expected square difference of distribution between two models

$$q(\epsilon) = E\left[(X_\epsilon(T_n) - Y_\epsilon(T_n))^2\right] \quad (28)$$

If $q(\epsilon)$ is minimized, the two models have the closest distributions.

Here $\epsilon = 0$ corresponds to ATM and both time dependent and time independent model trivially give the same distribution and $q(0) = 0$. The first derivative is shown to be zero at $\epsilon = 0$ below. In this circumstance, if second derivative of $q(\epsilon)$ is minimized there, the time independent model fits to time dependent model in the area close to ATM.

Now $q(\epsilon)$ is represented by

$$q(\epsilon) = E\left[\int_0^{T_n} \alpha_X^2(t)g_X^2(t)\left(h_\epsilon(t, X_\epsilon(t)) - \bar{h}_\epsilon(Y_\epsilon(t))\right)^2 dt\right] \quad (29)$$

Its first derivative is given by

$$q'(\epsilon) = 2E\left[\int_0^{T_n} \alpha_X^2(t)g_X^2(t)\left(h_\epsilon(t, X_\epsilon(t)) - \bar{h}_\epsilon(Y_\epsilon(t))\right)\left(\frac{\partial h_\epsilon(t, X_\epsilon(t))}{\partial \epsilon} - \frac{\partial \bar{h}_\epsilon(Y_\epsilon(t))}{\partial \epsilon}\right)dt\right] \quad (30)$$

And second derivative is

$$q''(\epsilon) = 2E\left[\int_0^{T_n} \alpha_X^2(t)g_X^2(t)\left(h_\epsilon(t, X_\epsilon(t)) - \bar{h}_\epsilon(Y_\epsilon(t))\right)\left(\frac{\partial^2 h_\epsilon(t, X_\epsilon(t))}{\partial \epsilon^2} - \frac{\partial^2 \bar{h}_\epsilon(Y_\epsilon(t))}{\partial \epsilon^2}\right)dt\right]$$
$$+ 2E\left[\int_0^{T_n} \alpha_X^2(t)g_X^2(t)\left(\frac{\partial h_\epsilon(t, X_\epsilon(t))}{\partial \epsilon} - \frac{\partial \bar{h}_\epsilon(Y_\epsilon(t))}{\partial \epsilon}\right)^2 dt\right] \quad (31)$$

It is apparent that $q(0) = q'(0) = 0$ and

$$q''(0) = 2E\left[\int_0^{T_n} \alpha_X^2(t)g_X^2(t)\left(\frac{\partial h_\epsilon(t, X_\epsilon(t))}{\partial \epsilon} - \frac{\partial \bar{h}_\epsilon(Y_\epsilon(t))}{\partial \epsilon}\right)^2_{\epsilon=0} dt\right] \quad (32)$$

We have

$$q''(0) = 2E\left[\int_0^{T_n} E\left[\alpha_X^2(t)(X_0(t) - X(0))^2\right]g_X^2(t)\left(\frac{\partial h(t, X(0))}{\partial x} - \frac{\partial \bar{h}(X(0))}{\partial x}\right)^2 dt\right]$$
$$= \int_0^{T_n} v^2(t)g_X^2(t)\left(\frac{\partial h(t, X(0))}{\partial x} - \frac{\partial \bar{h}b(X(0))}{\partial x}\right)^2 dt \quad (33)$$

Where

$$v^2(t) = E\left[\alpha_X^2(t)(X_0(t) - X(0))^2\right] \quad (34)$$

Therefore time independent local volatility, which minimizes $q''(0)$ is

$$\frac{\partial \bar{h}(X(0))}{\partial x} = \frac{\int_0^{T_n} v^2(t) g_X^2(t) \frac{\partial h(t, X(0))}{\partial x} dt}{\int_0^{T_n} v^2(t) g_X^2(t) dt} \quad (35)$$

Here for the CEV (and DD) type local volatility, derivative of a local volatility function by the rate at ATM is given by

$$\frac{\partial h(t, X(0))}{\partial x} = \frac{\beta_X(t)}{X(0)} \quad (36)$$

$$\frac{\partial \bar{h}(X(0))}{\partial x} = \frac{B}{X(0)} \quad (37)$$

Therefore skew averaging is given by

$$B = \frac{\int_0^{T_n} v^2(t) g_X^2(t) \beta_X(t) dt}{\int_0^{T_n} v^2(t) g_X^2(t) dt} \quad (38)$$

Now all we need to do is to estimate expectation value

$$v^2(t) = E\left[\alpha_X^2(t)\big(X_0(t) - X(0)\big)^2\right]$$

$$= E\left[E\left[\alpha_X^2(t)\big(X_0(t) - X(0)\big)^2 \mid \alpha_X()\right]\right]$$

$$= E\left[\int_0^t \alpha_X^2(u) g_X^2(u)\, \alpha_X^2(t)\right]$$

$$= \int_0^t E[\alpha_X^2(t) \alpha_X^2(u)]\, g_X^2(u) du$$

$$= \int_0^t \exp\left(5 \int_0^u v_X^2(s) ds + \int_0^t v_X^2(s) ds\right) g_X^2(u) du \quad (39)$$

Here $\alpha_X(0) = 1$ is used.

Estimation of $\alpha(0)$ and $\nu$

Let skew parameter of uncorrelated SABR $\beta$ be estimated already, we will derive $\alpha(0)$ and $\nu$ in this section [4].

We will choose $\alpha(0)$ and $\nu$ which approximate vanilla S waption prices of SABR/LMM. We will focus on the ATM path ($S(t) = S(0)$ and $L_i(t) = L_i(0)$ for $i = n, \cdots, m$). On ATM paths variance of Swap Rate from time 0 to T is given by

$$\int_0^{T_n} \lambda^2(t,\{L\}(0),\{\alpha\}(t))dt$$
$$= \sum_{i,j=n}^m v_i(0)v_j(0) \int_0^{T_n} \alpha_i(t)\alpha_j(t)g_i(t)g_j(t)\rho_{ij}(t)dt \quad (40)$$

Note that here we assume ATM path and the variance is not a function of rates but it is a function of stochastic volatility and it is a stochastic variable.

On the other hand, in Swap Rate uncorrelated SABR process, variance of the Swap Rate on ATM path is given by

$$\int_0^{T_n} \alpha^2(t)dt \quad (41)$$

Note that they are stochastic process also.

We will search $\alpha(0)$ and $v$ which give the same expected variance on ATM path. So we will solve

$$E\left[\sum_{i,j=n}^m v_i(0)v_j(0) \int_0^{T_n} \alpha_i(t)\alpha_j(t)g_i(t)g_j(t)\rho_{ij}(t)dt\right] = E\left[\int_0^{T_n} \alpha^2(t)dt\right] \quad (42)$$

It is possible to formulate the higher order equation but because it is too complex, it is not practical.

The R.H.S. is given by

$$E\left[\int_0^{T_n} \alpha^2(t)dt\right] = \frac{\alpha^2(0)}{v^2}\left(e^{T_n v^2} - 1\right) \quad (43)$$

And L.H.S. is given by

$$.E\left[\sum_{i,j=n}^m v_i(0)v_j(0) \int_0^{T_n} \alpha_i(t)\alpha_j(t)g_i(t)g_j(t)\rho_{ij}(t)dt\right]$$
$$= \sum_{i,j=n}^m v_i(0)v_j(0) \int_0^{T_n} \exp\left(r_{ij} \int_0^t v_i(u)v_j(u)du\right) g_i(t)g_j(t)\rho_{ij}(t)dt \quad (44)$$

Now we need to search $\alpha(0)$ and $v$ which satisfy

$$\frac{\alpha^2(0)}{v^2}\left(e^{T_n v^2} - 1\right) = \sum_{i,j=n}^m v_i(0)v_j(0) \int_0^{T_n} \exp\left(r_{ij} \int_0^t v_i(u)v_j(u)du\right) g_i(t)g_j(t)\rho_{ij}(t)dt \quad (45)$$

There are two variables to determine, and we need to introduce some constraint to solve it. Here we assume $\alpha(0)$ as

$$\alpha^2(0) = \left(\frac{1}{T_n}\right) \sum_{i,j=n}^{m} v_i(0) v_j(0) \int_0^{T_n} g_i(t) g_j(t) \rho_{ij}(t) dt \quad (46)$$

With this assumption, we can get ν by one of the following method

(1) Solve

$$\frac{\alpha^2(0)}{\nu^2}\left(e^{T_n \nu^2} - 1\right) = \sum_{i,j=n}^{m} v_i(0) v_j(0) \int_0^{T_n} \exp\left(r_{ij} \int_0^t v_i(u) v_j(u) du\right) g_i(t) g_j(t) \rho_{ij}(t) dt \quad (47)$$

Exactly.

(2) Assuming ν and $v_i(t)$ are small, expand the both side of the equation by ν. It gives

$$\nu^2 = \left(\frac{1}{T_n^2 \alpha^2(0)}\right) \sum_{i,j=n}^{m} v_i(0) v_j(0) \int_0^{T_n} \left\{r_{ij} \int_0^t v_i(u) v_j(u) du\right\} g_i(t) g_j(t) \rho_{ij}(t) dt \quad (48)$$

Summary of the approximate analytic formula

In this subsection, our analytic approximation of SABR/LMM introduced in this section, will be summarized. The approximation gives volatility of Swap SABR. The Time Dependent Skew parameter is given such that it gives the same slope of volatility surface on the ATM path. From the Time Dependent Skew, Time Independent Skew is given by the technique known as parameter (skew) averaging. Using these parameters, volatility of volatility is given such that the variance of Swap SABR is the same as the one from SABR/LMM. Based on the parameters of Swap SABR, exact solution of European Swaption price which is equivalent to the implied volatility is given.

[Fig 2]

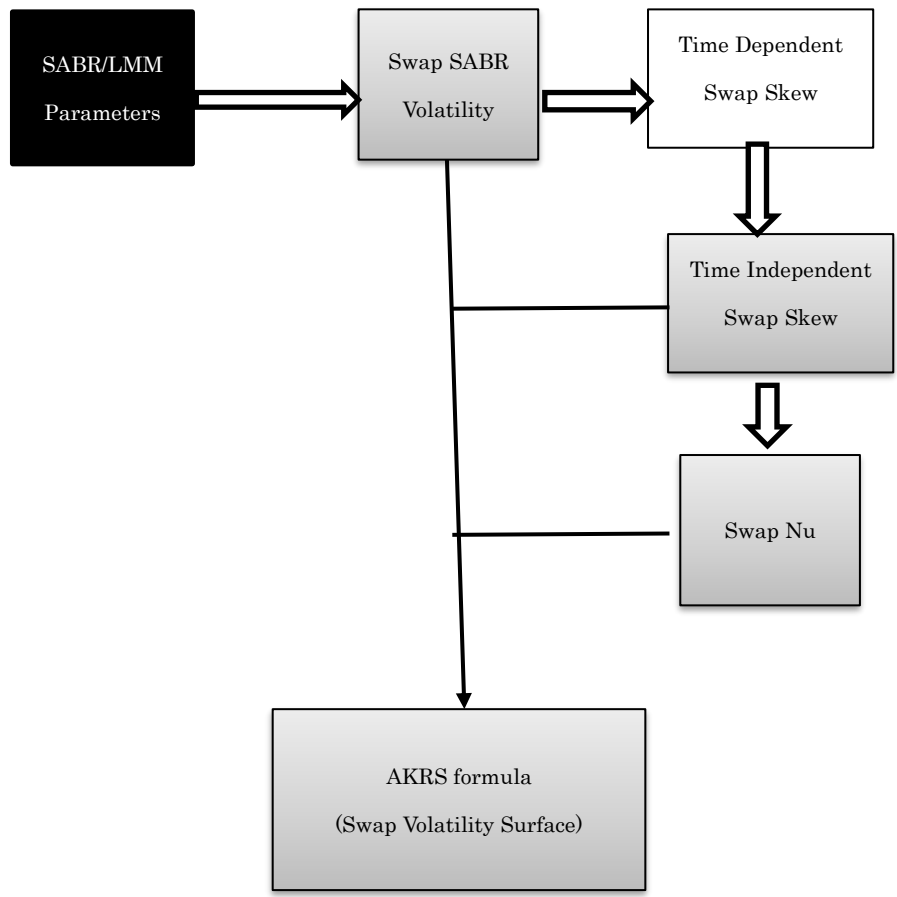

## Approximation formula for CMS spread (ATM)

The interest rate derivatives which depends on correlation between long and short rates directly is a (CMS) spread option. For the pricing of the callable derivatives which are related to the spread option such as callable CMS spread swap, the model should be ideally calibrated to the spread option market. So, in this subsection, the calibration formula for the spread option is given.

Approximation for pricing of spread option with a payoff $(S_1(t) - S_2(t) - K)^+$ will be given in this section. Here $S_1(t)$ and $S_2(t)$ are CMS rates with interest exchanges are at $T_{n+1}, \cdots, T_{n+a+1}$ and respectively at $T_{n+1}, \cdots, T_{n+b+1}$. Stochastic processes of Swap Rate underlying spread option are given by

$$dS_1(t) = \sum_{i=n}^{n+a} v_i^1(t)\alpha_i(t)\phi_i(t, L_i(t))g_i(t)dW_i(t) \quad (49)$$

$$dS_2(t) = \sum_{i=n}^{n+b} v_i^2(t)\alpha_i(t)\phi_i(t, L_i(t))g_i(t)dW_i(t) \quad (50)$$

We need to evaluate variance

$$E\left[\int_0^{T_n}(dS_1(t) - dS_2(t))^2\right] \quad (51)$$

Which is represented

$$E\left[\int_0^{T_n}(dS_1(t) - dS_2(t))^2\right]$$

$$= \sum_{i=n}^{n+a}\sum_{j=n}^{n+b}\int_0^{T_n} v_i^1(t)v_j^1(t)E[\alpha_i(t)\alpha_j(t)]\phi_i(t, L_i(t))\phi_j(t, L_j(t))g_i(t)g_j(t)\rho_{ij}(t)dt$$

$$- 2\sum_{i=n}^{n+a}\sum_{j=n}^{n+b}\int_0^{T_n} v_i^1(t)v_j^2(t)E[\alpha_i(t)\alpha_j(t)]\phi_i(t, L_i(t))\phi_j(t, L_j(t))g_i(t)g_j(t)\rho_{ij}(t)dt$$

$$+ \sum_{i=n}^{n+a}\sum_{j=n}^{n+b}\int_0^{T_n} v_i^2(t)v_j^2(t)E[\alpha_i(t)\alpha_j(t)]\phi_i(t, L_i(t))\phi_j(t, L_j(t))g_i(t)g_j(t)\rho_{ij}(t)dt \quad (52)$$

On the ATM path, this variance is evaluated as

$$E\left[\int_0^{T_n}(dS_1(t) - dS_2(t))^2\right]$$

$$= \sum_{i=n}^{n+a}\sum_{j=n}^{n+b}\int_0^{T_n} v_i^1(0)v_j^1(0)\exp\left(\int_0^t r_{ij}v_i(u)v_j(u)du\right)g_i(t)g_j(t)\rho_{ij}(t)dt$$

$$- 2\sum_{i=n}^{n+a}\sum_{j=n}^{n+b}\int_0^{T_n} v_i^1(t)v_j^2(t)(0)\exp\left(\int_0^t r_{ij}v_i(u)v_j(u)du\right)g_i(t)g_j(t)\rho_{ij}(t)dt$$

$$+ \sum_{i=n}^{n+a}\sum_{j=n}^{n+b}\int_0^{T_n} v_i^2(t)v_j^2(t)(0)\exp\left(\int_0^t r_{ij}v_i(u)v_j(u)du\right)g_i(t)g_j(t)\rho_{ij}(t)dt \quad (53)$$

## Calibration Scheme

The calibration scheme and parametrization of (interest rate) derivative models are based on the strategies of the banks. Sometimes the number of the model parameters is smaller than the market instruments for the model to be calibrated. In this case, optimization is necessary for the calibration. Here for the calculation of greeks, the bumping of market data can affect non-locally. It makes hedging rather difficult. So, in this paper, the number of market data is assumed to be the same as that of the model parameters as long as possible.

In this section, $S_{mn}(T_m)$ represents Swap Rate which starts at $T_m$ and has interest exchanges at $(T_{m+1}, \cdots, T_{n+1})$. The Zero SABR parameters which represents volatility surface of the swap is given by $\alpha_{mn}(0)$, $\beta_{mn}$, and $\nu_{mn}$.

## Bootstrapping

Ideally, our SABR/LMM should reproduce all of the SABR surfaces of the Swap Rates matrix. In this section, we will discuss bootstrapping methodology, which makes our model fit all of the Swap Rate SABR surfaces which are relevant for the pricing. In this calibration strategy, correlation parameters in the parametrization $\rho_{ij}(t) = \exp(-\beta(t)|T_i - T_j|)$ are calibrated in Global fit or input from outside.

All of our calibration process is based on the tenor structure we specified in the beginning of the paper. Before fitting SABR/LMM to market rate SABRs, we need interpolation which creates Swap Rate SABRs on our tenor structure.

The Swaption with underlying swap rate is $S_{nm}(T_n)$ and expiry date is $T_n$ depends in our approximation on volatilities of Libor $L_i(t)$ $(i = n, \cdots, m)$ with observation time $t = [0, T_n]$. Here SABR/LMM has three types of parameters $g_i(t)$, $v_i(t)$ and $\beta_i(t)$. Here we assume these parameters are assumed to be piecewise constant in observation time $t$ for tenor period $[T_i, T_{i+1}]$. The constant parameters for observation period $[T_i, T_{i+1}]$ are denoted as $g_i(l)$, $v_i(l)$ and $\beta_i(l)$. In the bootstrapping, the parameters are calibrated as $l: 0 \to N$ and for each $l$, $i: 0 \to N$. Here the target swaption is moving as $n: 0 \to N$, and for each $n$, as $m: 0 \to N$.

The bootstrapping method in this section is an ideal methodology. It is slow, and in most situations, it is unnecessarily complicated. And more serious drawback of this strategy is that it uses too much market data, and when our model fitted to all of them, sometimes the forward volatility of our model can become negative. Especially, the skew and smile, which don't have enough liquidity, suffer the situation more. It is because the market does not necessarily obey the assumption of our model.

So, we need a more stable and simple calibration strategy. In the next subsection, the Co-Terminal Calibration strategy, which is relevant for the pricing of Bermudan Swaption and other callable exotic derivatives.

### Co-Terminal Calibration Strategy

In the co-terminal calibration strategy, the model is calibrated to the volatility surfaces of the co-terminal swaptions with final maturity $T_{N+1}$. In the co-terminal calibration, volatility parameters $g_i(t) = g_i$, skew parameters $\beta_i(t) = \beta_i$ and smile parameter $\nu_i(t) = \nu_i$ are constant in observation time. The calibration scheme is the following

(1) Fit $g_i$ to $\alpha_{iN}(0)$
(2) Fit smile parameter $\nu_i$ to $\nu_{iN}$
(3) Fit skew parameter $\beta_i$ to $\beta_{iN}$

After the co-terminal calibration, time dependent volatility parameters $g_i(t)$ can be introduced and fitted to all the $\alpha_{mn}(0)$ to improce the fitting.

### Spread Option Calibration

In this subsection, ATM calibration to spread option market will be discussed. In the ATM calibration to spread option, only ATM volatility is fitted to the spread option market. Skew and smile is already calibrated to the whole volatility cube or co-terminal swaptions.

Let the lengths of underlying swaps of the spread option be $a$ and $b$, and $a < b$
The swap rate reset at date $T_n$, depend on Libors $L_i(t)$ with $i = n, \cdots, n + a$ and $i = n, \cdots, n + b$. Therefore the volatilities and correlation which affect the spread option price are $g_i(t)$ ($i = n, \cdots, n + b$). On the other hand, the spread option will be hedged by three (non-linear) products namely swaptions of $S_1(t)$ and $S_2(t)$ with option expiry $T_n$ and spread option between them. Spread option price is represented by normal volatility.

With this observation, we choose three parameters $g_n(T_n)$, $g_{n+b}(T_n)$ and $\beta(T_n)$ as free parameters which will be fitted in calibration.

Swaption SABR conversion

Our SABR/LMM does not have a correlation between Libor rates and stochastic volatilities. And our LMM calibrates to the uncorrelated SABR surface of the swaption market. On the other hand, in the swaption market, the volatility surfaces are represented by Hagan et al. approximation which corresponds to correlated SABR. So, before the calibration of our SABR/LMM, uncorrelated SABR needs to be calibrated to the Hagan et al. formula.

## Numerical Comparison

Now the approximation formula of Swap Rate SABR parameters in the SABR/LMM derived in the previous sections is compared to the numerical calculation of Monte Carlo simulation. Here implied volatilities of co-terminal European Swaption with final maturity 15 years are compared.

The outline of the results are as in the below graphs.

Fig 3

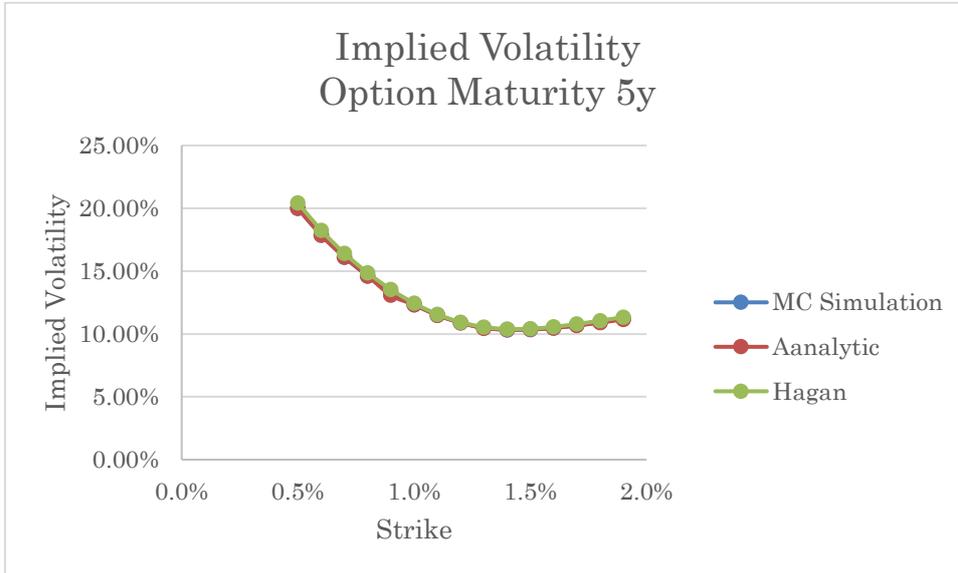

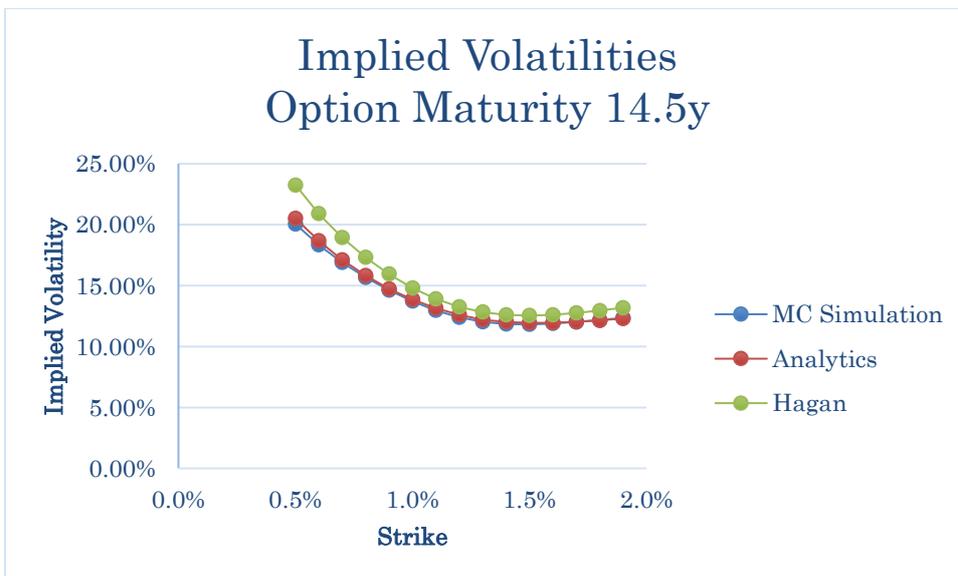

The details of the results of implied volatilities are given in the appendix. It is shown that the error in SABR/LMM is smaller than the error of the HKKW formula except for the short maturity swaptions where the Monte Carlo error is large in this setting.

The results on stressed market scenarios will be published elsewhere.

## Conclusion

The SABR/LMM with term structure of skew and smile is analyzed in detail. The approximate analytic function which maps parameters of SABR/LMM to parameters of swap rate SABR with zero correlation is given. Here the approximation is based on the combination of Ansatz for the volatility function, Parameter Averaging, and Variance Estimation for stochastic volatility. The approximation formula is compared to the Monte Carlo simulation for the pricing of European Swaption. The fitting between analytic approximation and Monte Carlo simulation is reasonably good.

There can be some extra research that is interesting and useful in practice. First, though we analyzed displaced diffusion SABR/LMM, we focused on the case of a positive rate environment. Since the great financial crisis of 2008, sometimes interest rates have been negative. To analyze the negative rate environment based on DD SABR/LMM would be useful.

The approximation formula for the CMS spread option for the ATM is derived. To incorporate skew of the spread option is also important.

## Acknowledgement

The author would like to thank Dr. Soetojo Tanudjaja for discussions and comments in this work.

## Appendix 1: AKRS formula

When the correlation between rate and volatility is zero, exact solution of the SABR model is given in the paper from Antonov, Konikov, Ruffino, and Spector [12]. In this appendix, the solution is briefly summarized.

In Swap rate SABR model, when correlations between swap rate and stochastic volatility is zero, the forward value of vanilla payers swaption is given by

$$V(0) = E[(S(T_n) - K)^+]$$
$$= E[E[(S(T_n) - K)^+|\alpha]]$$
$$= E\left[E\left[(S(T_n) - K)^+ | \int_0^{T_n} \alpha^2(u)du\right]\right] =$$
$$= E^\tau[C_{CEV}(\tau, K)] = \int_0^\infty d\tau \int_0^\infty p(T_n, v(T_n), \tau) C_{CEV}(\tau, K)$$

Where $C_{CEV}(\tau, K)$ is an call option price formula in CEV model, which depends on maturity and volatility only through $\tau = \int_0^{T_n} \alpha^2(u)du$, expectation $E^\tau[\ ]$ is the expectation about distribution of $\tau$ and $p(T_n, v(T_n), \tau)$ is a probability distribution function of $\tau$ with given $v(T_n)$. Combining representation of $p(T_n, v(T_n), \tau)$ and $C_{CEV}(\tau, K)$, forward price of payers swaption is given by

$$V(0) - (S(0) - K)^+$$
$$= \frac{2}{\pi}\sqrt{KS(0)} \left\{ \int_{s_-}^{s_+} ds \frac{\sin(\eta\phi(s))}{\sinh(s)} G(T_n v^2, s) + \sin(\eta\pi) \int_{s_+}^\infty ds \frac{e^{-\eta\psi(s)}}{\sinh(s)} G(T_n v^2, s) \right\}$$

Where

$$\eta = \frac{1}{|2(\beta - 1)|}$$

$$\phi(s) = 2\arctan\sqrt{\frac{\sinh^2(s) - \sinh^2(s_-)}{\sinh^2(s_+) - \sinh^2(s)}}$$

$$\psi(s) = 2\operatorname{arctanh}\sqrt{\frac{\sinh^2(s) - \sinh^2(s_+)}{\sinh^2(s) - \sinh^2(s_-)}}$$

$$s_- = \operatorname{arcsinh}\left(\frac{v|q - q_0|}{\alpha(0)}\right)$$

$$s_+ = \operatorname{arcsinh}\left(\frac{v|q + q_0|}{\alpha(0)}\right)$$

$$q = \frac{K^{1-\beta}}{1 - \beta}$$

$$q_0 = \frac{S(0)^{1-\beta}}{1-\beta}$$

$$G(t,s) = \frac{2\sqrt{2}e^{-\frac{t}{8}}}{t\sqrt{2\pi t}} \int_s^\infty du\, u e^{-u^2/2t} \sqrt{\cosh u - \cosh s}$$

## Appendix 2: Details in Numerical Calculation

The numerical calculation of SABR/LMM is done as following.

The calculation is done on the semi-annual tenor structure of 15 years. Based on the calibrated parameters of the SABR/LMM, calculate the parameters of uncorrelated SABR of co-terminal swaption with final maturity is 15 years.

The implied volatilities of co-terminal swaptions are compared for the following methodologies.
   1. Exact Solution of AKRS based on uncorrelated SABR parameters
   2. Hagan et al. approximation based on uncorrelated SABR parameters
   3. Monte Carlo simulation of SABR/LMM (CEV type local volatility)

In the Monte Carlo simulation, log normal Euler discretization is adopted. The tenor length is 6 months, and the simulation's time grid is monthly. The number of paths is 10,000. Regarding boundary conditions at forward Libor being zero, the Limited CEV scheme introduced in [2] is adopted.

Note that the Exact Solution in 1 is based on numerical integral, and there can be an error from it.

1. Exact Solution

| Maturity | Fwd | Strike | 0.5% | 0.6% | 0.7% | 0.8% | 0.9% | 1.0% | 1.1% | 1.2% | 1.3% | 1.4% | 1.5% | 1.6% | 1.7% | 1.8% | 1.9% |
|---|---|---|---|---|---|---|---|---|---|---|---|---|---|---|---|---|---|
| 0.5 | 0.013 | | 15.63% | 15.63% | 15.63% | 14.65% | 13.31% | 11.75% | 11.18% | 10.49% | 9.93% | 9.97% | 10.06% | 10.00% | 10.57% | 10.89% | 11.23% |
| 1 | 0.013 | | 20.51% | 18.22% | 16.30% | 14.66% | 13.28% | 12.12% | 11.19% | 10.52% | 10.01% | 10.00% | 10.08% | 10.01% | 10.56% | 10.87% | 11.19% |
| 1.5 | 0.013 | | 20.40% | 18.13% | 16.23% | 13.93% | 13.26% | 12.12% | 11.21% | 10.56% | 10.07% | 10.03% | 10.10% | 10.29% | 10.20% | 10.85% | 10.61% |
| 2 | 0.013 | | 20.29% | 18.04% | 16.17% | 14.59% | 13.24% | 12.13% | 11.23% | 10.59% | 10.12% | 10.06% | 10.12% | 10.29% | 10.54% | 10.83% | 11.13% |
| 2.5 | 0.013 | | 20.21% | 17.99% | 16.14% | 14.58% | 13.25% | 12.15% | 11.27% | 10.63% | 10.17% | 10.10% | 10.15% | 10.31% | 10.55% | 10.83% | 11.12% |
| 3 | 0.013 | | 20.14% | 17.94% | 16.11% | 14.56% | 12.86% | 12.16% | 11.30% | 10.67% | 10.22% | 10.14% | 10.18% | 10.33% | 10.28% | 10.82% | 11.11% |
| 3.5 | 0.013 | | 20.09% | 17.91% | 16.10% | 14.57% | 12.91% | 11.95% | 11.34% | 10.72% | 10.28% | 10.11% | 10.22% | 10.36% | 10.58% | 10.84% | 11.11% |
| 4 | 0.013 | | 20.04% | 17.88% | 16.09% | 14.57% | 13.29% | 12.23% | 11.39% | 10.77% | 10.33% | 10.24% | 10.26% | 10.39% | 10.60% | 10.85% | 11.12% |
| 4.5 | 0.013 | | 20.02% | 17.23% | 15.58% | 14.60% | 13.03% | 12.28% | 11.44% | 10.75% | 10.33% | 10.30% | 10.31% | 10.44% | 10.64% | 10.88% | 10.81% |
| 5 | 0.013 | | 19.99% | 17.87% | 16.11% | 14.62% | 13.09% | 12.32% | 11.50% | 10.89% | 10.46% | 10.35% | 10.36% | 10.48% | 10.67% | 10.90% | 11.15% |
| 5.5 | 0.013 | | 20.00% | 17.90% | 16.15% | 14.67% | 13.42% | 12.21% | 11.57% | 10.97% | 10.47% | 10.43% | 10.32% | 10.54% | 10.72% | 10.95% | 11.19% |
| 6 | 0.013 | | 20.00% | 17.92% | 16.18% | 14.41% | 12.29% | 11.53% | 10.97% | 10.55% | 10.50% | 10.49% | 10.60% | 10.78% | 10.78% | 11.23% | |
| 6.5 | 0.013 | | 20.00% | 17.93% | 15.86% | 14.76% | 13.53% | 12.52% | 11.71% | 11.12% | 10.63% | 10.57% | 10.47% | 10.66% | 10.83% | 11.04% | 11.27% |
| 7 | 0.013 | | 19.99% | 17.95% | 16.24% | 14.55% | 13.58% | 12.58% | 11.78% | 11.19% | 10.77% | 10.64% | 10.62% | 10.71% | 10.87% | 11.08% | 11.30% |
| 7.5 | 0.013 | | 20.02% | 18.00% | 16.30% | 14.87% | 13.66% | 12.66% | 11.87% | 11.28% | 10.86% | 10.73% | 10.71% | 10.79% | 10.95% | 11.15% | 11.37% |
| 8 | 0.013 | | 20.05% | 18.04% | 16.36% | 14.94% | 13.74% | 12.75% | 11.96% | 11.38% | 10.91% | 10.77% | 10.79% | 10.87% | 11.02% | 11.21% | 11.43% |
| 8.5 | 0.013 | | 20.08% | 18.08% | 16.42% | 15.01% | 13.82% | 12.83% | 12.05% | 11.47% | 11.05% | 10.91% | 10.81% | 10.86% | 10.98% | 11.28% | 11.48% |
| 9 | 0.013 | | 20.10% | 18.12% | 16.27% | 14.91% | 13.89% | 12.82% | 12.13% | 11.56% | 11.10% | 11.00% | 10.96% | 11.02% | 11.06% | 11.34% | 11.54% |
| 9.5 | 0.013 | | 20.12% | 18.15% | 16.51% | 15.13% | 13.96% | 12.99% | 12.22% | 11.60% | 11.23% | 11.04% | 11.04% | 11.10% | 11.14% | 11.30% | 11.48% |
| 10 | 0.013 | | 20.14% | 18.18% | 16.56% | 15.18% | 14.02% | 13.06% | 12.30% | 11.73% | 11.31% | 11.16% | 11.11% | 11.17% | 11.29% | 11.37% | 11.65% |
| 10.5 | 0.013 | | 20.02% | 18.24% | 16.50% | 15.16% | 14.11% | 13.16% | 12.40% | 11.83% | 11.42% | 11.23% | 11.21% | 11.26% | 11.38% | 11.54% | 11.63% |
| 11 | 0.013 | | 20.23% | 18.29% | 16.69% | 15.34% | 14.19% | 13.25% | 12.49% | 11.93% | 11.52% | 11.36% | 11.30% | 11.34% | 11.46% | 11.61% | 11.79% |
| 11.5 | 0.013 | | 20.27% | 18.33% | 16.75% | 15.41% | 14.27% | 13.34% | 12.59% | 12.03% | 11.61% | 11.45% | 11.39% | 11.43% | 11.54% | 11.69% | 11.86% |
| 12 | 0.013 | | 20.19% | 18.38% | 16.80% | 15.47% | 14.35% | 13.42% | 12.68% | 12.12% | 11.71% | 11.52% | 11.48% | 11.51% | 11.61% | 11.76% | 11.93% |
| 12.5 | 0.013 | | 20.36% | 18.45% | 16.27% | 14.91% | 14.44% | 13.52% | 12.78% | 12.23% | 11.81% | 11.65% | 11.58% | 11.61% | 11.70% | 11.84% | 12.01% |
| 13 | 0.013 | | 20.40% | 18.45% | 16.94% | 15.64% | 14.53% | 13.61% | 12.88% | 12.33% | 11.92% | 11.74% | 11.68% | 11.70% | 11.79% | 11.92% | 12.08% |
| 13.5 | 0.013 | | 20.45% | 18.58% | 16.96% | 15.67% | 14.61% | 13.68% | 12.98% | 12.42% | 12.02% | 11.85% | 11.77% | 11.77% | 11.84% | 11.96% | 12.16% |
| 14 | 0.013 | | 20.49% | 18.63% | 17.07% | 15.78% | 14.69% | 13.79% | 13.07% | 12.53% | 12.12% | 11.94% | 11.86% | 11.88% | 11.96% | 12.08% | 12.23% |
| 14.5 | 0.013 | | 20.53% | 18.69% | 17.13% | 15.85% | 14.75% | 13.88% | 13.16% | 12.62% | 12.21% | 12.04% | 11.95% | 11.96% | 12.03% | 12.15% | 12.30% |

2. HKKW formula

| Maturity | Strike | 0.5% | 0.6% | 0.7% | 0.8% | 0.9% | 1.0% | 1.1% | 1.2% | 1.3% | 1.4% | 1.5% | 1.6% | 1.7% | 1.8% | 1.9% |
|---|---|---|---|---|---|---|---|---|---|---|---|---|---|---|---|---|
| 0.5 | | 20.59% | 18.31% | 16.38% | 14.74% | 13.33% | 12.15% | 11.20% | 10.52% | 10.12% | 10.00% | 10.08% | 10.30% | 10.59% | 10.92% | 11.26% |
| 1 | | 20.50% | 18.25% | 16.34% | 14.71% | 13.32% | 12.15% | 11.22% | 10.54% | 10.15% | 10.02% | 10.10% | 10.31% | 10.59% | 10.91% | 11.24% |
| 1.5 | | 20.45% | 18.21% | 16.32% | 14.70% | 13.32% | 12.17% | 11.24% | 10.58% | 10.18% | 10.05% | 10.12% | 10.32% | 10.60% | 10.91% | 11.23% |
| 2 | | 20.39% | 18.17% | 16.29% | 14.69% | 13.32% | 12.18% | 11.27% | 10.61% | 10.22% | 10.08% | 10.14% | 10.34% | 10.60% | 10.90% | 11.22% |
| 2.5 | | 20.37% | 18.16% | 16.29% | 14.70% | 13.34% | 12.21% | 11.31% | 10.65% | 10.26% | 10.13% | 10.18% | 10.36% | 10.62% | 10.92% | 11.22% |
| 3 | | 20.35% | 18.14% | 16.29% | 14.71% | 13.36% | 12.24% | 11.34% | 10.69% | 10.31% | 10.17% | 10.21% | 10.39% | 10.64% | 10.93% | 11.23% |
| 3.5 | | 20.36% | 18.16% | 16.31% | 14.74% | 13.40% | 12.28% | 11.39% | 10.75% | 10.36% | 10.22% | 10.26% | 10.43% | 10.67% | 10.95% | 11.25% |
| 4 | | 20.36% | 18.17% | 16.33% | 14.77% | 13.44% | 12.33% | 11.44% | 10.80% | 10.42% | 10.27% | 10.31% | 10.47% | 10.70% | 10.98% | 11.27% |
| 4.5 | | 20.40% | 18.22% | 16.38% | 14.82% | 13.49% | 12.39% | 11.51% | 10.87% | 10.49% | 10.33% | 10.37% | 10.52% | 10.75% | 11.02% | 11.31% |
| 5 | | 20.44% | 18.25% | 16.42% | 14.87% | 13.55% | 12.45% | 11.57% | 10.94% | 10.55% | 10.40% | 10.42% | 10.57% | 10.80% | 11.06% | 11.34% |
| 5.5 | | 20.51% | 18.33% | 16.50% | 14.94% | 13.62% | 12.53% | 11.65% | 11.02% | 10.64% | 10.48% | 10.50% | 10.64% | 10.86% | 11.12% | 11.40% |
| 6 | | 20.59% | 18.41% | 16.57% | 15.02% | 13.70% | 12.61% | 11.74% | 11.11% | 10.72% | 10.56% | 10.57% | 10.71% | 10.93% | 11.18% | 11.46% |
| 6.5 | | 20.66% | 18.48% | 16.65% | 15.09% | 13.78% | 12.69% | 11.82% | 11.19% | 10.80% | 10.64% | 10.65% | 10.78% | 10.99% | 11.24% | 11.51% |
| 7 | | 20.73% | 18.55% | 16.72% | 15.17% | 13.86% | 12.77% | 11.90% | 11.27% | 10.88% | 10.71% | 10.72% | 10.85% | 11.05% | 11.30% | 11.57% |
| 7.5 | | 20.87% | 18.67% | 16.84% | 15.28% | 13.97% | 12.88% | 12.01% | 11.38% | 10.99% | 10.82% | 10.82% | 10.95% | 11.15% | 11.39% | 11.65% |
| 8 | | 20.99% | 18.80% | 16.96% | 15.40% | 14.08% | 12.99% | 12.12% | 11.49% | 11.10% | 10.92% | 10.92% | 11.04% | 11.24% | 11.48% | 11.74% |
| 8.5 | | 21.12% | 18.92% | 17.07% | 15.51% | 14.19% | 13.10% | 12.23% | 11.60% | 11.21% | 11.03% | 11.02% | 11.14% | 11.33% | 11.57% | 11.82% |
| 9 | | 21.25% | 19.04% | 17.19% | 15.63% | 14.31% | 13.21% | 12.35% | 11.71% | 11.32% | 11.13% | 11.12% | 11.23% | 11.42% | 11.65% | 11.91% |
| 9.5 | | 21.38% | 19.16% | 17.31% | 15.74% | 14.42% | 13.32% | 12.46% | 11.82% | 11.43% | 11.24% | 11.23% | 11.33% | 11.51% | 11.74% | 11.99% |
| 10 | | 21.51% | 19.28% | 17.43% | 15.86% | 14.53% | 13.44% | 12.57% | 11.93% | 11.53% | 11.34% | 11.33% | 11.43% | 11.61% | 11.83% | 12.08% |
| 10.5 | | 21.68% | 19.45% | 17.58% | 16.00% | 14.67% | 13.58% | 12.71% | 12.07% | 11.67% | 11.47% | 11.45% | 11.55% | 11.72% | 11.94% | 12.19% |
| 11 | | 21.86% | 19.61% | 17.73% | 16.15% | 14.82% | 13.72% | 12.84% | 12.21% | 11.80% | 11.60% | 11.57% | 11.67% | 11.84% | 12.06% | 12.30% |
| 11.5 | | 22.03% | 19.77% | 17.89% | 16.30% | 14.96% | 13.86% | 12.98% | 12.34% | 11.93% | 11.73% | 11.70% | 11.79% | 11.96% | 12.17% | 12.41% |
| 12 | | 22.21% | 19.94% | 18.04% | 16.45% | 15.11% | 14.00% | 13.12% | 12.48% | 12.07% | 11.86% | 11.83% | 11.91% | 12.07% | 12.28% | 12.52% |
| 12.5 | | 22.41% | 20.13% | 18.23% | 16.62% | 15.27% | 14.16% | 13.28% | 12.64% | 12.22% | 12.01% | 11.97% | 12.05% | 12.21% | 12.42% | 12.65% |
| 13 | | 22.62% | 20.33% | 18.41% | 16.80% | 15.44% | 14.33% | 13.44% | 12.79% | 12.37% | 12.16% | 12.12% | 12.19% | 12.35% | 12.55% | 12.79% |
| 13.5 | | 22.83% | 20.52% | 18.60% | 16.98% | 15.61% | 14.49% | 13.60% | 12.95% | 12.53% | 12.31% | 12.26% | 12.33% | 12.49% | 12.69% | 12.92% |
| 14 | | 23.05% | 20.72% | 18.78% | 17.15% | 15.79% | 14.66% | 13.77% | 13.11% | 12.68% | 12.46% | 12.41% | 12.48% | 12.62% | 12.82% | 13.05% |
| 14.5 | | 23.26% | 20.92% | 18.97% | 17.33% | 15.96% | 14.83% | 13.93% | 13.27% | 12.84% | 12.61% | 12.56% | 12.62% | 12.76% | 12.96% | 13.18% |

3. Monte Carlo simulation of SABR/LMM

| Maturity | Strike | 0.5% | 0.6% | 0.7% | 0.8% | 0.9% | 1.0% | 1.1% | 1.2% | 1.3% | 1.4% | 1.5% | 1.6% | 1.7% | 1.8% | 1.9% |
|---|---|---|---|---|---|---|---|---|---|---|---|---|---|---|---|---|
|  | 0 |  |  |  |  |  |  |  |  |  |  |  |  |  |  |  |
| 0.5 | 1 | 43.47% | 35.55% | 28.83% | 22.98% | 17.76% | 13.34% | 11.09% | 10.40% | 10.15% | 10.04% | 10.13% | 10.58% | 9.89% | 7.81% | 7.81% |
| 1 | 2 | 31.88% | 26.08% | 21.16% | 17.15% | 14.45% | 12.58% | 11.33% | 10.57% | 10.20% | 10.07% | 10.20% | 10.38% | 10.65% | 10.88% | 10.87% |
| 1.5 | 3 | 31.27% | 25.72% | 21.16% | 17.38% | 14.49% | 12.58% | 11.49% | 10.71% | 10.30% | 10.19% | 10.29% | 10.51% | 10.78% | 10.94% | 11.07% |
| 2 | 4 | 25.65% | 21.42% | 18.05% | 15.62% | 13.76% | 12.40% | 11.38% | 10.68% | 10.28% | 10.12% | 10.20% | 10.43% | 10.69% | 10.88% | 10.98% |
| 2.5 | 5 | 28.05% | 23.38% | 19.68% | 16.82% | 14.59% | 12.93% | 11.71% | 10.88% | 10.44% | 10.31% | 10.40% | 10.63% | 10.90% | 11.19% | 11.44% |
| 3 | 6 | 24.92% | 21.07% | 18.06% | 15.75% | 13.95% | 12.57% | 11.52% | 10.81% | 10.41% | 10.26% | 10.30% | 10.48% | 10.70% | 10.93% | 11.11% |
| 3.5 | 7 | 24.24% | 20.66% | 17.85% | 15.70% | 13.98% | 12.65% | 11.63% | 10.92% | 10.51% | 10.32% | 10.32% | 10.48% | 10.75% | 11.05% | 11.38% |
| 4 | 8 | 21.19% | 18.66% | 16.68% | 15.01% | 13.62% | 12.49% | 11.58% | 10.94% | 10.55% | 10.34% | 10.35% | 10.47% | 10.67% | 10.90% | 11.17% |
| 4.5 | 9 | 19.83% | 17.85% | 16.18% | 14.75% | 13.50% | 12.39% | 11.51% | 10.90% | 10.51% | 10.32% | 10.34% | 10.48% | 10.69% | 10.91% | 11.17% |
| 5 | 10 | 19.99% | 17.88% | 16.12% | 14.62% | 13.37% | 12.33% | 11.50% | 10.90% | 10.49% | 10.33% | 10.36% | 10.47% | 10.69% | 10.94% | 11.23% |
| 5.5 | 11 | 20.59% | 18.24% | 16.34% | 14.80% | 13.48% | 12.41% | 11.58% | 10.98% | 10.59% | 10.41% | 10.43% | 10.56% | 10.75% | 11.01% | 11.31% |
| 6 | 12 | 19.85% | 17.82% | 16.10% | 14.62% | 13.36% | 12.34% | 11.52% | 10.94% | 10.58% | 10.44% | 10.46% | 10.60% | 10.80% | 11.03% | 11.28% |
| 6.5 | 13 | 19.86% | 17.87% | 16.15% | 14.67% | 13.43% | 12.42% | 11.62% | 11.03% | 10.67% | 10.53% | 10.55% | 10.66% | 10.86% | 11.10% | 11.34% |
| 7 | 14 | 19.53% | 17.65% | 16.07% | 14.70% | 13.52% | 12.53% | 11.75% | 11.18% | 10.82% | 10.66% | 10.66% | 10.76% | 10.93% | 11.14% | 11.38% |
| 7.5 | 15 | 19.34% | 17.59% | 16.07% | 14.76% | 13.61% | 12.63% | 11.84% | 11.28% | 10.92% | 10.76% | 10.78% | 10.89% | 11.08% | 11.31% | 11.56% |
| 8 | 16 | 19.30% | 17.59% | 16.09% | 14.79% | 13.64% | 12.67% | 11.88% | 11.32% | 10.96% | 10.80% | 10.79% | 10.90% | 11.08% | 11.31% | 11.55% |
| 8.5 | 17 | 19.17% | 17.49% | 16.03% | 14.76% | 13.65% | 12.70% | 11.93% | 11.36% | 11.00% | 10.83% | 10.82% | 10.93% | 11.09% | 11.29% | 11.52% |
| 9 | 18 | 20.00% | 18.07% | 16.44% | 15.05% | 13.86% | 12.85% | 12.04% | 11.45% | 11.07% | 10.90% | 10.87% | 10.96% | 11.11% | 11.31% | 11.54% |
| 9.5 | 19 | 19.93% | 18.05% | 16.44% | 15.07% | 13.88% | 12.87% | 12.04% | 11.46% | 11.08% | 10.91% | 10.89% | 10.96% | 11.11% | 11.33% | 11.56% |
| 10 | 20 | 19.80% | 17.94% | 16.35% | 14.98% | 13.81% | 12.83% | 12.03% | 11.47% | 11.11% | 10.94% | 10.90% | 10.98% | 11.14% | 11.34% | 11.58% |
| 10.5 | 21 | 19.91% | 18.06% | 16.49% | 15.12% | 13.96% | 12.99% | 12.20% | 11.64% | 11.27% | 11.11% | 11.08% | 11.16% | 11.30% | 11.49% | 11.70% |
| 11 | 22 | 19.46% | 17.75% | 16.26% | 14.96% | 13.85% | 12.92% | 12.14% | 11.59% | 11.24% | 11.08% | 11.06% | 11.13% | 11.28% | 11.47% | 11.66% |
| 11.5 | 23 | 19.95% | 18.14% | 16.58% | 15.25% | 14.09% | 13.13% | 12.36% | 11.80% | 11.44% | 11.27% | 11.25% | 11.33% | 11.47% | 11.64% | 11.85% |
| 12 | 24 | 20.07% | 18.26% | 16.73% | 15.39% | 14.23% | 13.26% | 12.50% | 11.95% | 11.58% | 11.41% | 11.38% | 11.46% | 11.60% | 11.78% | 11.98% |
| 12.5 | 25 | 20.35% | 18.50% | 16.92% | 15.56% | 14.39% | 13.41% | 12.63% | 12.05% | 11.67% | 11.49% | 11.46% | 11.55% | 11.69% | 11.87% | 12.06% |
| 13 | 26 | 20.28% | 18.44% | 16.88% | 15.55% | 14.40% | 13.45% | 12.70% | 12.13% | 11.75% | 11.57% | 11.55% | 11.62% | 11.77% | 11.94% | 12.13% |
| 13.5 | 27 | 20.00% | 18.25% | 16.76% | 15.49% | 14.39% | 13.47% | 12.71% | 12.14% | 11.77% | 11.60% | 11.57% | 11.64% | 11.77% | 11.93% | 12.12% |
| 14 | 28 | 19.84% | 18.15% | 16.73% | 15.51% | 14.46% | 13.56% | 12.83% | 12.27% | 11.90% | 11.71% | 11.67% | 11.74% | 11.88% | 12.06% | 12.27% |
| 14.5 | 29 | 20.06% | 18.35% | 16.91% | 15.69% | 14.64% | 13.73% | 12.97% | 12.42% | 12.04% | 11.86% | 11.82% | 11.90% | 12.02% | 12.18% | 12.35% |